\documentstyle[osa,manuscript]{revtex}  % DON'T CHANGE

\begin{document}                % INITIALIZE - DONT CHANGE
\title{Spherical Structures in the Inflationary Cosmology}
\author{N. Riazi}
\address{Physics Department and Biruni Observatory, Shiraz University,\\
Shiraz 71454, Iran,\\and\\
Institute for Research in Physics and Mathematics (IPM),\\
Farmanieh, Tehran, Iran. \\ email: riazi@sun01.susc.ac.ir}
\maketitle
\begin{abstract}                % DON'T CHANGE THIS LINE
It has been suggested that wormholes and other non-trivial geometrical
structures might have been formed during the quantum cosmological
era ($t\sim 10^{-43}$s). Subsequent inflation of the universe might
have enlarged these structures to macroscopic sizes. 
In this paper, spherical geometrical structures in an inflationary RW
background are derived from the Einstein equations, using a
constraint on the energy-momentum tensor which is an extension of
the one expected for inflation. The possibility  of dynamical wormholes
and other spherical structures  are explored
in the framework of the solutions.
\end{abstract}
\section{Introduction}               % Introduction goes below.
It is now generally accepted that the universe underwent a brief
inflationary period at $t\sim 10^{-35}$s, during which the
scale factor of the universe increased exponentially (or quasi-exponentially).
Such a dynamics emerges as a result of an equation of state which
resembles that of the vacuum or cosmological constant \cite{gu1}.
It has been suggested that non-trivial topological structures probably
formed during the quantum cosmological era ($t\sim 10^{-43}$s).
Among these structures are wormholes which are throat-like objects
connecting two otherwise disconnected spacetimes or two remote parts
of the same spacetime. 
Wormholes, originally of the Planck length scale,
are expected to take part in the cosmological
expansion during the inflationary era, and turn into macroscopic objects
\cite{fr1,ro1}.
Sato et al.\cite{sa1} investigated the possibility
of wormhole formation during the inflationary era.
Roman\cite{ro1} presented a new classical metric which
represents a wormhole, embedded in a flat de Sitter spacetime.
The throat radius was shown to inflate
with time.

Static wormholes are known to  violate the
weak and strong energy conditions.
Even the averaged weak energy conditions are known to be violated
by static wormholes\cite{ha1,ti1,bo1,mo1}.
Morris and Thorne\cite{mo1} discussed the conditions that a wormhole
must satisfy in order to be traversable in both directions.
These conditions are derived from the requirement that there
are no event horizons or curvature singularities. Since the
energy-momentum which is required to support
static, traversable wormholes does not satisfy the null
(and the weak) energy conditions, they called the corresponding
matter "exotic". Teo\cite{te1} discussed the general form of a
stationary, axially symmetric traversable wormhole. He found
that the null energy condition is violated by such wormholes,
but there can be classes of traversing geodesics which do
not cross energy condition violating regions. 
Evolving wormholes, on the other
hand, may alter this unfavorable situation\cite{ka1,ri1}.
Although the violation of the energy conditions is unacceptable
from the conventional point of view, it has been shown that
some quantum field theory effects (such as the Casimir effect) allow
such a violation\cite{ep1}. 
Moreover, the negative
energy densities required for the support of the wormhole geometry
might be produced by the gravitational squeezing of the vacuum\cite{ho1}.
The possibility of closed, timelike curves in wormhole geometries
is another concern in the wormhole theory\cite{ha2}.
The  role of the cosmological constant (i.e. the vacuum energy-momentum tensor)
on the dynamics and stability of the Lorentzian wormholes was studied
by Kim\cite{ki1}.

There are two approaches in formulating wormhole spacetimes: 1) By joining
two asymptotically flat (or asymptotically RW) spacetimes via a
boundary layer with $\delta$-function energy-momentum\cite{vi1},
and 2) By smoothly merging the wormhole metrics to a cosmological background
\cite{ro1,ri1}.  In the present paper, we employ
the latter method to explore the dynamics of spherical structures
(including wormholes) which join smoothly to a cosmological
background in its inflationary phase.  In section 2, we formulate the
problem and seek various exact solutions which correspond to different
choices of the integration constants. Section 3 is devoted to the
corresponding energy-momentum tensor and the exoticity parameter.
The last section contains a summary and a few  concluding remarks.
\section{Spherical solutions in an inflationary background}
It is well-known that the most general isotropic and homogeneous metric consistent
with the cosmological principle is the Robertson-Walker (RW) metric
\begin{equation}
\label{rwmet}
ds^2=-dt^2+R^2(t)\left[ \frac{dr^2}{1-kr^2}+r^2 d\theta^2 +r^2 \sin^2 \theta d\phi^2\right],
\end{equation}
in which $(r,\theta ,\phi )$ are the so-called co-moving coordinates, $R(t)$ is
the scale factor, and $k=0,\pm 1$ correspond to the
flat, closed, and open universes, respectively.
This metric has 6 Killing vectors\cite{we1}, representing its high
degree of symmetry (Minkowski spacetime has 10 Killing vectors).

It is clear that the existence of a spherical object in a RW background spoils
part of its symmetries. The resulting spacetime is no longer  homogeneous. Accordingly,
we start by assuming the evolving, spherically symmetric spacetime
\begin{equation}
\label{sphmet}
ds^2=-dt^2+R^2(t)\left[ \left( 1+a(r)\right) dr^2+r^2 d\theta^2 +r^2 \sin^2 \theta d\phi^2\right],
\end{equation}
where $a(r)$ is an unknown function. It is clear that the RW metric (\ref{rwmet}) is a special
case of (\ref{sphmet}), with $1+a(r)=\frac{1}{1-kr^2}$. It can be shown  that
in general, the
metric (\ref{sphmet}) possesses the following Killing vectors,
with respect to the
$(\frac{\partial}{\partial t},\frac{\partial}{\partial r},
\frac{\partial}{\partial \theta},\frac{\partial}{\partial \phi})$
basis\cite{na1}:
\[
{\bf K_1}=(1,0,0,0),
\]
\[
{\bf K_2}=(0,0,\cos \phi, -\cot \theta \sin \phi ),
\]
\[
{\bf K_3}=(0,0,\sin \phi ,\cot \theta \cos \phi ),
\]
\begin{equation}
{\bf K_4}=(0,0,0,1),
\end{equation}

The non-vanishing components of the Einstein tensor for
the metric (\ref{sphmet}) read
\begin{equation}
G_{tt} =-3\frac{\dot{R}^2}{R^2}-\frac{ra'+a+a^2}{r^2R^2(1+a)^2},
\end{equation}
\begin{equation}
G_{rr}=(1+a)(\dot{R}^2+2R\ddot{R})+\frac{a}{r^2},
\end{equation}
and
\begin{equation}
G_{\theta\theta}=2r^2R\ddot{R}+r^2\dot{R}^2+\frac{ra'}{2(1+a)^2}
=\frac{1}{\sin^2 \theta}G_{\phi\phi}.
\end{equation}
Such a spacetime is supported by an anisotropic energy-momentum tensor:
\begin{equation}
\label{rho}
\rho (r,t)=-\frac{1}{8\pi G}T^t_t=\frac{1}{8\pi G}\left[
\frac{3\dot{R}^2}{R^2}+\frac{ra'+a+a^2}{r^2R^2(1+a)^2}\right] ,
\end{equation}
\begin{equation}
\label{pr}
P_r (r,t)=\frac{1}{8\pi G}T^r_r =-\frac{1}{8\pi G}
\left[ \frac{\dot{R}^2}{R^2}+
\frac{2\ddot{R}}{R}+\frac{a}{r^2R^2(1+a)}\right] ,
\end{equation}
and
\begin{equation}
\label{pt}
P_t (r,t)=\frac{1}{8\pi G}T^\theta_\theta =-\frac{1}{8\pi G}\left[
\frac{\dot{R}^2}{R^2}+
\frac{2\ddot{R}}{R}+\frac{a'}{2rR^2(1+a)^2}\right] .
\end{equation}
In these equations, $P_r$ and $P_t$ are the radial and transverse pressures,
and "dots" and "primes" denote differentiation
with respect to $r$ and $t$, respectively. It can be easily shown that
the special case $P_r=P_t$ leads to either
$a=0$ or $a(r)=\frac{r^2}{r_o^2-r^2}$,
which upon substituting  in
(\ref{rho}), lead to the disappearance of
$r$ dependence in $\rho (r,t)$ as expected for a homogeneous spacetime.

Since we are looking for spherical structures in an inflationary background, $P_r$ and
$P_t$ should have the asymptotic behavior
\begin{equation}
P_r\rightarrow P_t \rightarrow -\rho \rightarrow {\rm constant},
\end{equation}
far from the central structure.
Recall that for vacuum (i.e. the cosmological constant),
$P_r=P_t=-\rho ={\rm constant}$.
In such a case, the background spacetime is expected
to expand exponentially (i.e. $R(t)\propto \exp (Ht)$, with $H$ constant).
As a generalization of the vacuum equation of state, we consider the relation
\begin{equation}
\label{cond}
\rho =-\frac{1}{1+\gamma}\left[ P_r +\gamma P_t\right],
\end{equation}
in which $\gamma$ is a constant.
Using (\ref{rho})-(\ref{pt}), one obtains from
(\ref{cond})
\[
3\frac{\dot{R}^2}{R^2}+\frac{ra'+a(1+a)}{r^2R^2(1+a)^2}=
\]
\begin{equation}
\frac{1}{1+\gamma}\left\{ \frac{\dot{R}^2}{R^2}+\frac{2\ddot{R}}{R}+
\frac{a}{r^2R^2(1+a)}+\gamma \frac{2\ddot{R}}{R}+\gamma \frac{\dot{R}^2}{R^2}
+\gamma \frac{a'}{2rR^2(1+a)^2}\right\}.
\end{equation}

This equation can be separated into t-dependent and r-dependent parts:
\begin{equation}
2\dot{R}^2-2\ddot{R}R=\frac{-(1+\gamma )\left[ ra'+a(1+a)\right] +
a(1+a)+\gamma  ra' /2}{(1+\gamma )r^2(1+a)^2}.
\end{equation}
The lhs and rhs parts, should therefore be constant, independent of r and t. Let us
call the constant of separation $2c_1$. We first seek solutions for the special case
$c_1=0$.
\subsection{Solutions for the case $c_1=0$}
The equation $\dot{R}^2=\ddot{R}R$ is immediately solved to obtain
\begin{equation}
R(t)=R_o e^{Ht},
\end{equation}
in which $R_o$ and $H$ are constants. The equation for $a(r)$ reads
\begin{equation}
ra'=- \frac{2\gamma }{2+\gamma }a(1+a),
\end{equation}
which can be integrated to obtain the exact solution
\begin{equation}
\label{speca}
a(r)=\frac{r_o^\alpha}{r^\alpha -r_o^\alpha},
\end{equation}
where $\alpha=\frac{2\gamma}{2+\gamma}$, and $r_o$ is the constant of integration,
having the same dimensions as $r$. The resulting metric is
\begin{equation}
ds^2=-dt^2 +R_o^2 e^{2Ht}\left[ \frac{r^\alpha}{r^\alpha-r_o^\alpha}dr^2
+r^2d\theta^2 +r^2\sin^2 \theta d\phi^2\right].
\end{equation}
We therefore have a wormhole-centered asymptotically de Sitter spacetime, with
the central wormhole having a throat circumference
\begin{equation}
\ell =R(t)r_o \oint d\phi =2\pi r_o R_o e^{Ht},
\end{equation}
which expands with time, exponentially.

The energy momentum tensor needed to support this structure has the components
\begin{equation}
\rho (r,t)=\frac{3H^2}{8\pi G}\left[ 1+\left( \frac{2-\gamma}{2+\gamma}\right)
\frac{r_o^{\frac{2\gamma}{2+\gamma}}}{3H^2R_o^2r^{\frac{4(1+\gamma)}{2
+\gamma}}}e^{-2Ht} \right],
\end{equation}
\begin{equation}
P_r (r,t)=-\frac{3H^2}{8\pi G}\left[ 1+
\frac{r_o^{\frac{2\gamma}{2+\gamma}}}{3H^2R_o^2r^{\frac{4(1+\gamma)}{2
+\gamma}}}e^{-2Ht} \right],
\end{equation}
and
\begin{equation}
P_t (r,t)=-\frac{3H^2}{8\pi G}\left[ 1- \left( \frac{\gamma}{2+\gamma}\right)
\frac{r_o^{\frac{2\gamma}{2+\gamma}}}{3H^2R_o^2r^{\frac{4(1+\gamma)}{2
+\gamma}}}e^{-2Ht} \right].
\end{equation}
It is seen that for reasonable values of $\gamma$, the equation of state
rapidly approaches that of the cosmological constant  for $r>>r_o$
and/or $t>>1/H$. Moreover, for the special case $\gamma =2$, the energy
density is homogeneous, independent of both $r$ and $t$. It can be easily
shown that this result is consistent with the energy equation
\begin{equation}
D_\mu T^\mu_\nu =0,
\end{equation}
which leads to
\begin{equation}
\frac{\partial \rho}{\partial t}+\left[ 3\rho +P_r+2P_t\right] =0.
\end{equation}
Note that $3\rho +P_r +2P_t =0$ in the $\gamma =2$ case.
\subsection{The general case $c_1 \neq 0$}
Let us proceed with the general case $c_1\neq 0$;
\begin{equation}
\dot{R}^2-R\ddot{R}=c_1.
\end{equation}
This equation can be integrated once to obtain
\begin{equation}
\frac{R^2}{R_o^2}=\dot{R}^2-c_1,
\end{equation}
where $R_o$ is a constant of integration. Integrating once again,
leads to
\begin{equation}
t=\int \frac{dR}{\left[ c_1+\frac{R^2}{R_o^2}\right]^{1/2}},
\end{equation}
or
\begin{equation}
R(t)=R_o \sinh (Ht), \ \ \ {\rm for} \ \ \ c_1>0,
\end{equation}
and
\begin{equation}
R(t)=R_0 \cosh (Ht), \ \ \ {\rm for } \ \ \ c_1<0,
\end{equation}
where $H=\frac{\sqrt{|c_1|}}{R_o}$.

The radial equation  reads
\begin{equation}
\label{gena}
\frac{da(r)}{dr}=Ar(1+a)^2-B\frac{a(1+a)}{r},
\end{equation}
where
\begin{equation}
A=-\frac{4c_1(1+\gamma)}{2+\gamma},\ \ \ {\rm and}\ \ \ B=\frac{2\gamma}{2+\gamma}.
\end{equation}
Fortunately, equation (\ref{gena}) can also be solved exactly to obtain
\begin{equation}
\label{genaa}
1+a(r)=\left[1+ c_or^{-B}-\frac{A}{2+B}r^2  \right]^{-1},
\end{equation}
in which $c_o$ is a constant of integration. Note that for the special case
$c_1=0$, we have $A=0$, and the solution (\ref{genaa}) reduces to
(\ref{speca}), with a suitable definition for $c_o$. The function $a(r)+1$ is
plotted against $r$ in Figure 1, for $c_1=\pm 1$,
$c_o=\pm 1$, and $\gamma =2$. Note that for $\gamma =2$, 
(\ref{genaa}) can be written as
\begin{equation}
\label{solar}
1+a (r)=\frac{r}{c_1 r^3+r+c_o}=\frac{r}{(r-r_1)(c_1r^2+c_1r_1r-\frac{c_o}{r_1})},
\end{equation}
where
\begin{equation}
\label{wormrad}
r_1= \beta -\frac{1}{3c_1\beta},
\end{equation}
and
\begin{equation}
\beta=-\frac{c_o}{2c_1}+\frac{\sqrt{3}}{18c_1}\sqrt{\frac{4+27c_o^2 c_1}{c_1}}.
\end{equation}
In order for $r_1$ to be real, we must have
\begin{equation}
c_1 >0,
\end{equation}
or
\begin{equation}
c_1<0, \ \ \ {\rm and}\ \ \ c_0^2>\frac{4}{27|c_1|}.
\end{equation}

The choice $c_1=\pm 1$ and $c_0=0$ leads to $1+a(r)=\frac{1}{1\pm r^2}$, which
correspond to the $k=\mp 1$ RW metric. Referring to Figure 1, we
see that for $c_1=-1$ and $c_0=-1$, $1+a(r)$ becomes
negative and the metric has an improper signature. Such a solution,
therefore, is not of cosmological significance.

The $c_1=1$ and $c_0=1$ case is a metric which is regular
everywhere except at $r=0$.  The $c_1=-1$ and $c_0 =1$  case
possesses an upper bound on $r$ ($=r_1$) and corresponds to
a closed cosmological background.

The $c_1=1$ and $c_0=-1$ and similar cases possess a lower bound at $r=r_1$,
and correspond to wormhole-centered open universes.
Note that in such cases, $r_1$ plays the role of the
wormhole radius, and we also have
\begin{equation}
1+a(r)\rightarrow \frac{1}{r^2}, \ \ \ {\rm as} \ \ \ r\rightarrow \infty ,
\end{equation}
in accordance with the $k=-1$ RW model. The function $1+a(r)$ for the
closed and open RW universes are also shown in Figure 1 for comparison.

Note that in the $c_1=c_0=1$ case, there is a singularity at $r=0$,
where the diagonal components of the Einstein tensor diverge.
       \section{Energy-momentum tensor and exoticity parameter}
Using equations (\ref{rho}) to (\ref{pt}), it can be shown that for the general
case $c_1\neq 0$, we have
\begin{equation}
\rho =\frac{1}{8\pi G}\left[ \frac{3}{R_o^2}+\frac{f_1(r)}{R^2}\right] ,
\end{equation}
where
\begin{equation}
f_1(r)=3c_1+A+\frac{(1+B)a}{(1+a)r^2},
\end{equation}
\begin{equation}
P_r=-\frac{1}{8\pi G}\left[ \frac{3}{R_o^2}+\frac{f_2(r)}{R^2} \right] ,
\end{equation}
where
\begin{equation}
f_2(r)=c_1+\frac{a}{(1+a)r^2},
\end{equation}
and
\begin{equation}
P_t=-\frac{1}{8\pi G}\left[ \frac{3}{R_o^2}+\frac{f_3(r)}{R^2}\right] ,
\end{equation}
where
\begin{equation}
f_3(r)=c_1+\frac{1}{2}A(1+a)+\frac{1}{2}\frac{Ba}{r^2}.
\end{equation}
The explicit form of the function $a(r)$ is given by (\ref{genaa}).
The exoticity parameter is defined according to
\begin{equation}
\label{exo}
\xi =-\frac{P+\rho}{|\rho|},
\end{equation}
with $\xi >0$ corresponding to the exotic matter and $\xi <0$ to the
non-exotic. Since we have an unisotropic medium, we modify (\ref{exo})
according to
\begin{equation}
\label{modexo}
\bar{\xi}=-\frac{\bar{P}+\rho}{|\rho|},
\end{equation}
where
\begin{equation}
\bar{P}=\frac{1}{1+\gamma}(P_r+\gamma P_t).
\end{equation}
Note that for the vacuum equation of state we have $\xi =0$, which is the boarder line
between the exotic and non-exotic matter. From (\ref{cond}), we
see that the modified exoticity parameter (\ref{modexo}) vanishes
everywhere and for all cases.

Let us check the null energy condition
\begin{equation}
R_{\mu\nu}k^\mu k^\nu \ge 0,
\end{equation}
where $R_{\mu\nu}$ is the Ricci tensor corresponding to the metric
(\ref{sphmet}), and $k^\mu$ is a null vector taken to be\cite{vi1}
\begin{equation}
k^\mu =(\sqrt{-g^{tt}}, \pm \sqrt{g^{rr}},0,0).
\end{equation}
Using the metric (\ref{sphmet}), we readily obtain
\begin{equation}
R_{\mu\nu}k^\mu k^\nu=2\frac{\ddot{R}}{R}-2\frac{\dot{R}^2}{R^2}
-\frac{a'}{R^2r(1+a)^2}.
\end{equation}
Using the solutions obtained in the previous section for the
$c_1=0$, $c_1>0$, and $c_1<0$ cases, we obtain
\begin{equation}
R_{\mu\nu}k^\mu k^\nu =
{\rm cosech}^2 Ht\left[ \frac{Bc_0}{R_o^2r^{2+B}}
-\frac{2A}{(2+B)R_o^2}-2H^2\right];\ \ \ {\rm for }\ \ \  c_1>0,
\end{equation}
\begin{equation}
R_{\mu\nu}k^\mu k^\nu =
\frac{2\gamma}{2+\gamma}\frac{1}{R_0^2r^2}e^{-2Ht}\left(
\frac{r_o}{r}\right)^\alpha;\ \ \ {\rm for}\ \ \  c_1=0,
\end{equation}
and
\begin{equation}
R_{\mu\nu}k^\mu k^\nu =
{\rm sech}^2 Ht\left[ 2H^2 +\frac{Bc_o}{R_0^2r^{2+B}}
-\frac{2A}{(2+B)R_o^2}\right]; \ \ \ {\rm for} \ \ \  c_1<0.  
\end{equation}
It is seen that the matter everywhere satisfies the null energy condition
for the $c_1=0$ case. For the $c_1>0$ and $c_0<0$ case which contains
wormhole solutions, the null energy condition reduces to
\begin{equation}
\frac{8c_1(1+\gamma )}{(2+\gamma)(2+B)R_0^2}\ge 2H^2+\frac{B|c_0|}{R_0^2
r_1^{2+B}},
\end{equation}
where $r_1$ is the wormhole throat radius given by (\ref{wormrad}).
\section{Summary and conclusion}
If wormholes are produced spontaneously in the quantum era,
as claimed by some quantum cosmology considerations,
we would expect them to undergo a rapid expansion during
the inflationary era, and turn into macroscopic objects.
Motivated by this interesting possibility, we studied the
dynamics of a spherical geometrical structure in an inflationary
background. 
Based on a reasonable constraint on the energy-momentum tensor
of an anisotropic medium reminiscent of the vacuum, we solved
the Einstein equations to obtain a metric which  smoothly
merges into  a rapidly expanding background
universe. Solutions were classified into different categories,
with distinct geometries for the central object.
One class of the solutions was shown to represent Lorentzian
wormholes with the throat radius expanding with the universe.

The present work differs from Roman\cite{ro1} in
the following respects. In the ansatz metric used by Roman,
the scale factor is pre-assumed to be of
the form $e^{2\chi t}$ with $2\chi$ corresponding to
the expansion rate ($H$ in the present paper).
The scale factor is {\it derived} here, with  our
$c_1=0$ case coinciding with the one adopted by Roman
with $\Phi (r)=0$ ($\Phi$ is the redshift function).
Moreover, while the shape function is left undetermined in
the mentioned reference, it is explicitly (and analytically)
calculated here.  
      
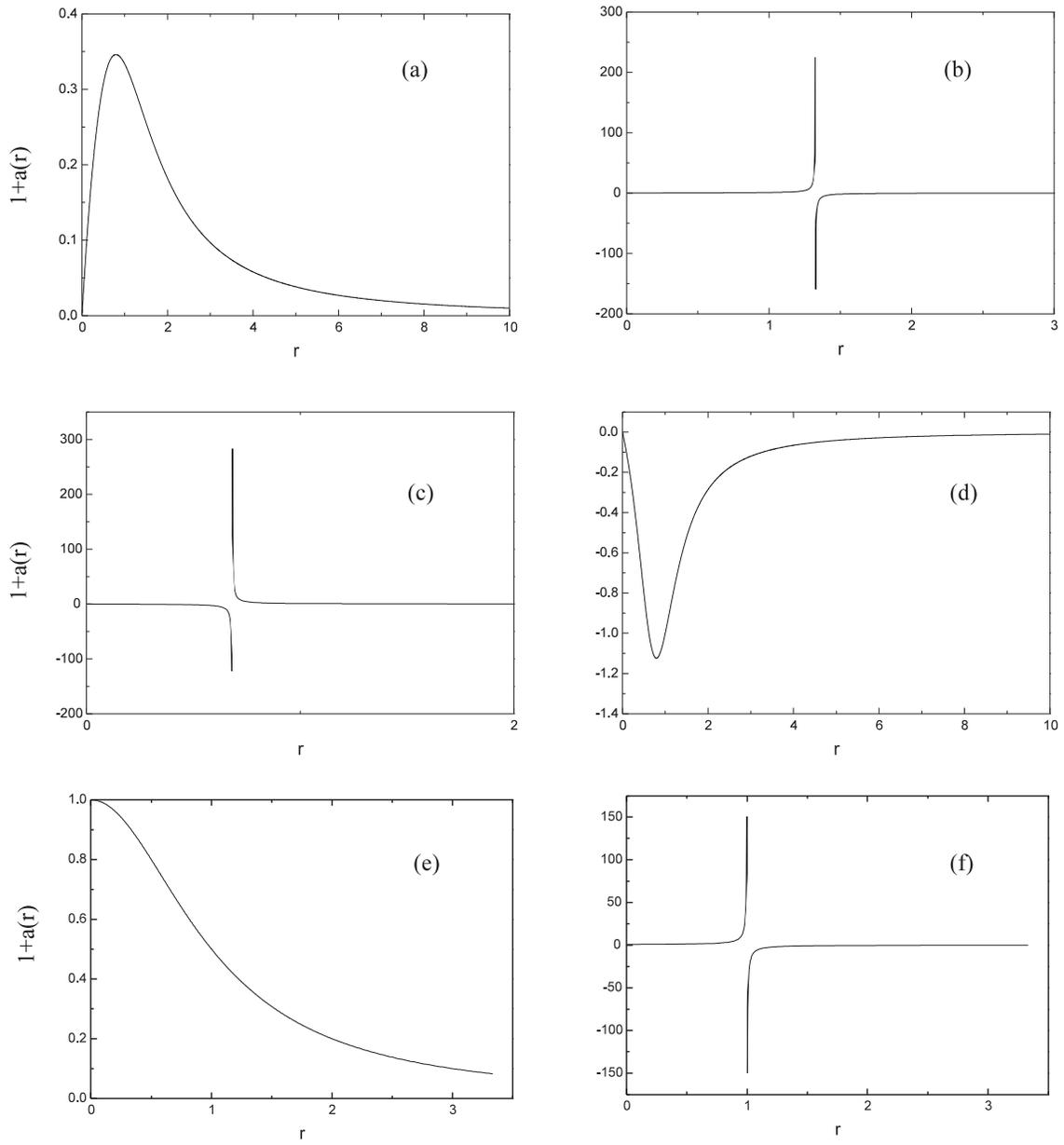
\begin{figure}  % Please send figures with disk, or separately if
% if it is an e-mail submission. (Good photo or India ink drawing.)
\caption{The radial function $1+a(r)$ for (a) $c_1=1$, $c_0=1$,
(b)  $c_1=-1$, $c_0=1$, (c) $c_1=1$, $c_0=-1$, 
and (d)  $c_1=-1$, $c_0=-1$. The radial function $1+a(r)$ for the
Robertson-Walker  metric is also shown for comparison: (e) open RW with
$k=-1$, and (f) closed RW with $k=+1$.}
\end{figure}

\end{document}